\documentclass[journal,twocolumn,twoside]{IEEEtran} 

\makeatletter
\let\NAT@parse\undefined
\makeatother

\usepackage{amsmath,amssymb,amsfonts}
\usepackage[final]{graphicx}

\usepackage{cite}

\usepackage{ucs} 
\usepackage[utf8x]{inputenc}

\usepackage[usenames,dvipsnames]{xcolor}
\usepackage{latexsym}
\usepackage[nolist,nohyperlinks]{acronym}
\usepackage{multirow}
\usepackage{multicol}

\newtheorem{corollary}{Corollary}

\newtheorem{lemma}{Lemma}

\newtheorem{remark}{Remark}

\def \L {{\mathcal L }}

\begin{acronym}
\acro{FTR}{fluctuating two-ray}
\acro{IFTR}{independent fluctuating two-ray}
\acro{LOS}{line of sight}
\acro{NLOS}{non line of sight}
\acro{PDF}{probability density function}
\acro{CDF}{cumulative density function}
\acro{CCDF}{complementary \ac{CDF}}
\acro{MGF}{moment generating function}
\acro{TWDP}{two-wave with diffuse power}
\acro{SNR}{signal-to-noise ratio}
\acro{AWGN}{additive white Gaussian noise}
\acro{BER}{bit error rate}
\acro{CEP}{conditional error probability}
\acro{LMS}{land mobile satellite}
\acro{mmWave}{millimeter-wave}
\acro{UAC}{underwater acoustic communications}
\acro{KS}{Kolmogorov-Smirnov}
\acro{5G}{fifth generation}
\acro{6G}{sixth generation}
\end{acronym}

\makeatletter
\def\blfootnote{\xdef\@thefnmark{}\@footnotetext}
\makeatother

\begin{document}
\title{The Fluctuating Two-Ray Fading Model with Independent Specular Components}

\vspace{3mm}
\author{ Maryam Olyaee, Jos\'e A. Cort\'es, F. Javier Lopez-Martinez, Jos\'e F. Paris and Juan M. Romero-Jerez
\thanks{\noindent The authors are with Communications and Signal Processing Lab, Instituto Universitario de Investigaci\'on en Telecomunicaci\'on (TELMA), Universidad de M\'alaga, CEI Andaluc\'ia TECH.
ETSI Telecomunicaci\'on, Bulevar Louis Pasteur 35, 29010, M\'alaga, Spain. F.J. Lopez Martinez is also with the Department of Signal Theory, Networking and Communications, Universidad de Granada, 18071, Granada, Spain.(e-mails: maryam.olyaee@ic.uma.es, jaca@ic.uma.es, paris@ic.uma.es, fjlopezm@ic.uma.es, romero@dte.uma.es).
\\
\indent This work has been funded in part by the Spanish Government and the European Fund for
Regional Development FEDER (project TEC2017-87913-R), by Junta de Andaluc\'ia (projects P18-RT-3175 and P18-RT-3587) and by Universidad de M\'alaga (project UMA20-FEDERJA-002).
}}

\maketitle
\begin{abstract}
We introduce and characterize the independent fluctuating two-ray (IFTR) fading model, a class of fading models consisting of two specular components which fluctuate \textit{independently}, plus a diffuse component modeled as a complex Gaussian random variable. The IFTR model complements the popular fluctuating two-ray (FTR) model, on which the specular components are fully correlated and fluctuate \textit{jointly}. The chief probability functions of the received SNR in IFTR fading, including the PDF, CDF and MGF, are expressed in closed-form, having a functional form similar to other state-of-the-art fading models. Then, the IFTR model is empirically validated using multiple channels measured in rather diverse scenarios, including line of sight (LOS) millimeter-wave, land mobile satellite (LMS) and underwater acoustic communication (UAC), showing a better fit than the original FTR model and other models previously used in these environments. Additionally, the performance of wireless communication systems operating under IFTR fading is evaluated in closed-form in two scenarios: (\textit{i}) exact and asymptotic bit error rate for a family of coherent modulations; and (\textit{ii}) exact and asymptotic outage probability. 

\end{abstract}

\vspace{0mm}

\IEEEpeerreviewmaketitle

\section{Introduction}
\label{intro}


\IEEEPARstart{S}{mall-scale} fading is a key propagation effect in many wireless scenarios. The accurate modeling of this phenomenon in the \ac{mmWave} band has become particularly relevant because of its use in the \ac{5G} standard. This also applies to \ac{LMS} channels, which are attracting an increasing interest as the \ac{6G} standard is expected to integrate space-ground links to provide global coverage \cite{Wang00}. This propagation phenomenon also occurs even in environments where information is conveyed by means of non-electromagnetic waves, as in \ac{UAC} \cite{Canete16}. 

The Rice distribution has been largely employed to model \ac{LOS} wireless channels. Recent attempts to characterize fading in \ac{mmWave} channels, for instance, have used it \cite{Samimi2016, Sun17}. Two refinements of the Ricie model are of particular interest because of their relevant physical interpretation: one considers that the \ac{LOS} (or specular) component considered in the Rice model fluctuates randomly. This yields the Rician-shadowed distribution, which has been employed to model \ac{LMS} channels \cite{Abdi2003}. The second one considers an additional specular component, yielding the \ac{TWDP} model \cite{Durgin2002,Rao2015}, which has proven to outperform the Rice distribution when modeling indoor channels at 60 GHz \cite{Zochmann19}.

The \ac{FTR} model was recently proposed as a natural generalization of \textit{both} the Rician-shadowed and the \ac{TWDP} models, and many others \cite{Romero2017}. It consists of two specular components with random phases that jointly fluctuate, plus a diffuse component. The \ac{FTR} model not only provides a better fit to \ac{mmWave} field measurements than previous fading models, but also its primary probability functions - \ac{CDF}, \ac{PDF} and \ac{MGF} - can be expressed in closed-form. 

However, the physical model that originates FTR fading assumes that the two specular components experience a common (i.e., fully correlated) fluctuation, which may not always be the case in practice. For instance, the specular components may fluctuate because of two different phenomena as they reach the destination through different paths, thus being affected by different scatterers and/or perturbations. As we will later see, this situation is quite common in practice. 
Besides capturing the reality of a number of propagation mechanisms, allowing the specular components to fluctuate \textit{independently} provides an additional degree of freedom that results in an improved modeling of these scenarios, which can be quite significant in some cases. For instance, in terms of the \ac{CDF}, this is translated into an increased flexibility to change its log-log concavity and convexity, yielding a better fit to experimental data. 

A special case of this scenario was originally suggested in \cite{Romero2017}, where the two specular components undergo independent and identically distributed (iid) Nakagami-$m$ fluctuations. However, a thorough statistical characterization and analysis was not presented, and has never been carried out in the literature to the best of our knowledge. In this paper, we introduce, characterize and validate the \ac{IFTR} fading model to complement the \ac{FTR} model by allowing the specular components to fluctuate independently and experience dissimilar fading severity, i.e., the specular components fluctuate according to non-necessarily identical distributions. Specifically, the following key contributions are made in this context:
\begin{itemize}
\item Closed-form expressions for the \ac{PDF}, \ac{CDF}, and \ac{MGF} of the received power envelope, or equivalently the \ac{SNR}, under \ac{IFTR} fading are obtained.
\item It is shown that the \ac{IFTR} model fits measured channels in rather diverse scenarios such as \ac{mmWave}, \ac{LMS} and \ac{UAC} better than the \ac{FTR} model and other state-of-the-art models previously used in these environments. 
\item The performance of communication systems operating in these propagation conditions is analyzed, in terms of the exact and asymptotic \ac{BER} for a family of modulations, and in terms of the outage probability.
\item The influence of the model parameters on the fading statistics and on the system performance is investigated, and the key differences with respect to the FTR model are assessed.
\end{itemize}

The remainder of this paper is organized as follows. Section II presents the physical channel model for the \ac{IFTR} fading. Its statistical characterization in terms of the \ac{MGF}, \ac{PDF}, and \ac{CDF} of the received \ac{SNR} is carried out in Section III. Section IV shows the empirical validation of the newly proposed IFTR fading model. Based on the obtained statistical functions, performance analysis of wireless communications systems undergoing \ac{IFTR} fading is exemplified in Section V. Numerical results are given in Section VI, and the main conclusions are summarized in Section VII.

\section{Channel model}
\label{systemmodel}

Let us assume that the small-scale fluctuations in the amplitude of a signal transmitted over a wireless channel are given by two fluctuating dominant waves, referred to as specular components, to which other diffusely propagating waves are added. The complex baseband voltage of the wireless channel of this model can be expressed as 
\begin{equation}
\label{eq:01}
\begin{split}
V_r  = \sqrt {\zeta _1 } V_1  \exp \left( {j\phi _1 } \right) +  \sqrt {\zeta _2 } V_2 \exp \left( {j\phi _2 } \right) + X + jY,
\end{split}
\end{equation}
where $ \sqrt {\zeta _i } V_i \exp \left( {j\phi _i } \right)$ represents the \emph{i-th} specular component $(i=1,2)$, which is assumed to have an average amplitude $V_i$ modulated by a random variable $\sqrt {\zeta _i }$ responsible for its fluctuation, where $\zeta_1$ and $\zeta_2$ are independent unit-mean Gamma distributed random variables with \ac{PDF}
\begin{equation}
\label{eq:02}
f_{\zeta _i} \left( u \right) = \frac{m_i ^{m_i } u^{m_i  - 1}}{\Gamma \left( {m_i } \right)} e^{ - m_i u} ,\quad \quad i = 1,2.
\end{equation}
Without any loss of generality, in the sequel we will assume $V_1\geq V_2$. The \emph{i-th} specular component is assumed to have a uniformly distributed random phase $\phi _i $, such that $\phi _i \sim \mathcal{U}[0,2\pi)$, with $\phi _1 $ and $\phi _2 $ considered to be statistically independent. 
On the other hand, $X + jY$ is a complex Gaussian random variable, such that $X,Y \sim \mathcal{N}(0,\sigma^2)$, representing the diffuse received signal component due to the combined reception of numerous weak, independently-phased scattered waves. This model will be referred to as the \ac{IFTR} fading model.
Note that if $m_1,m_2 \rightarrow \infty$ then $\zeta_1,\zeta_2 \rightarrow 1$, i.e., the fluctuation of the specular components tends to disappear and the \ac{IFTR} model tends to the \ac{TWDP} fading model proposed by Durgin, Rappaport and De Wolf \cite{Durgin2002}, also known as the Generalized Two-Ray fading model with Uniformly distributed phases (GTR-U) \cite{Rao2015}. 

As with the TWDP fading model, the \ac{IFTR} channel model can conveniently be described in terms of the parameters $K$ and $\Delta$, defined as
\begin{equation}
\label{eq:03}
K \triangleq \frac{V_1^2+V_2^2}{2\sigma^2},
\end{equation}
\begin{equation}
\label{eq:04}
\Delta \triangleq \frac{2V_1 V_2}{V_1^2+V_2^2},
\end{equation}
where the parameter $K$ represents the ratio of the average power of the dominant (specular) components to the power
of the remaining diffuse multipath. On the other hand, $\Delta$ is a parameter ranging from 0 to 1 capturing how similar to each other are the average received powers of the specular components: when the average magnitudes of the two specular components are equal, we have $\Delta=1$. Conversely, in the absence of a second component
($V_1=0$ or $V_2=0$), then $\Delta=0$. The \ac{IFTR} model is an \textit{alternative} generalization of the \ac{TWDP} model, which differs from the original FTR model proposed by the authors in \cite{Romero2017}. The IFTR model can be applied in those situations in which the specular components follow very different paths and are affected by different scatterers.

\section{Statistical characterization}
\label{characterization}

We will first characterize the distribution of the received power envelope associated with the \ac{IFTR} fading model, or equivalently, the distribution of the received \ac{SNR}. After passing through the multipath fading channel, the signal will be affected by additive white Gaussian noise (AWGN) with one-sided power spectral density $N_0$. The statistical characterization of the instantaneous \ac{SNR}, here denoted as $\gamma$, is crucial for the analysis and design of wireless communications systems, as many performance metrics in wireless communications are a function of the \ac{SNR}.

The received average \ac{SNR} $\bar\gamma$ after transmitting a symbol with energy density $E_s$ undergoing a multipath fading channel as described in (\ref{eq:01}) will be
\begin{equation}
\begin{split}
\label{eq:05}
  & \bar \gamma  = \left( {E_s /N_0 } \right){\mathbb E}\left\{ {\left| {V_r } \right|^2 } \right\} = \left( {E_s /N_0 } \right)\left( {V_1^2  + V_2^2  + 2\sigma ^2 } \right)  \cr 
  & \quad \quad  = \left( {E_s /N_0 } \right)2\sigma ^2 \left( {1 + K} \right), \cr
\end{split}
\end{equation}
where $\mathbb E\{\cdot\}$ denotes the expectation operator.

With all the above definitions, the chief probability functions related to the \ac{FTR} fading model can now be computed.

\subsection{MGF}
In the following lemma we show that, for the \ac{IFTR} fading model, it is possible to obtain the \ac{MGF} of $\gamma$ in closed-form.

\begin{lemma}  
Let us consider the \ac{IFTR} fading model as described in (\ref{eq:01})-(\ref{eq:02}). Then, the \ac{MGF} of the received \ac{SNR} $\gamma$ will be given by (\ref{eq:06}), where $_2F_1 (\cdot)$ is the Gauss hypergeometric function \cite[p. 556 (15.1.1)]{Abramowitz70}.

\begin{figure*}[!t]
\normalsize
\begin{equation} \label{eq:06}
\begin{split}
  & M_\gamma  \left( s \right) = {{1 + K} \over {1 + K - \bar \gamma s}}{{m_1^{m_1 } } \over {\left[ {m_1  - {K \over 2}\left( {1 + \sqrt {1 - \Delta ^2 } } \right){{\bar \gamma s} \over {1 + K - \bar \gamma s}}} \right]^{m_1 } }}{{m_2^{m_2 } } \over {\left[ {m_2  - {K \over 2}\left( {1 - \sqrt {1 - \Delta ^2 } } \right){{\bar \gamma s} \over {1 + K - \bar \gamma s}}} \right]^{m_2 } }}  \cr 
  & \quad \quad  \times \,_2 F_1 \left( {m_1 ,m_2 ;1;{{K^2 \Delta ^2 } \over {\left[ {2m_1 {{1 + K - \bar \gamma s} \over {\bar \gamma s}} - K\left( {1 + \sqrt {1 - \Delta ^2 } } \right)} \right]\left[ {2m_2 {{1 + K - \bar \gamma s} \over {\bar \gamma s}} - K\left( {1 - \sqrt {1 - \Delta ^2 } } \right)} \right]}}} \right). \cr
\end{split}
\end{equation}
\hrulefill
\vspace*{4pt}
\end{figure*}

\end{lemma}

\begin{IEEEproof} See Appendix \ref{App1}.
\end{IEEEproof}

When any of the parameters $m_1$ or $m_2$ takes an integer value, the \ac{MGF} of the \ac{SNR} in the \ac{IFTR} fading model can be calculated as a finite sum of elementary functions. 

\begin{corollary}  
When $m_1\in\mathbb{Z}^+$, the \ac{MGF} of the \ac{SNR} $\gamma$ of the \ac{IFTR} fading channel can be expressed as a finite sum of elementary terms as given in (\ref{eq:100}).

\begin{figure*}[!t]
\normalsize
\begin{equation} \label{eq:100}
\begin{split}
  & M_\gamma  \left( s \right) = \frac{{1 + K}}
{{1 + K - \bar \gamma s}}m_1^{m_1 } m_2^{m_2 } \left[ {m_1  - \frac{K}
{2}\left( {1 + \sqrt {1 - \Delta ^2 } } \right)\frac{{\bar \gamma s}}
{{1 + K - \bar \gamma s}}} \right]^{m_2  - m_1 } \sum\limits_{n = 0}^{m_1  - 1} {\frac{1}
{{n!}}\binom{m_1-1}{n}} \frac{{\Gamma \left( {m_2  + n} \right)}}
{{\Gamma \left( {m_2 } \right)}}  \cr 
  & \quad  \times \left( {\frac{{K\Delta }}
{2}\frac{{\bar \gamma s}}
{{1 + K - \bar \gamma s}}} \right)^{2n} \left[ {m_1 m_2  - \left( {m_1 \frac{K}
{2}\left( {1 - \sqrt {1 - \Delta ^2 } } \right) + m_2 \frac{K}
{2}\left( {1 + \sqrt {1 - \Delta ^2 } } \right)} \right)\frac{{\bar \gamma s}}
{{1 + K - \bar \gamma s}}} \right]^{ - m_2  - n}. \cr
\end{split}
\end{equation}
\hrulefill
\vspace*{4pt}
\end{figure*}

\end{corollary}

\begin{IEEEproof} See Appendix \ref{App2}.
\end{IEEEproof}

\begin{remark} 
For the case when $m_2\in\mathbb{Z}^+$ and $m_1$ is not necessarily an integer, the \ac{MGF} of $\gamma$ can be calculated by using (\ref{eq:100}) and interchanging $m_1$ and $m_2$ and also interchanging the occurences of ${ + \sqrt {1 - \Delta ^2 } }$
and ${ - \sqrt {1 - \Delta ^2 } }$. In the sequel, the interchanging of these parameters will also hold in the subsequent expressions obtained for $m_1\in\mathbb{Z}^+$, when the case $m_2\in\mathbb{Z}^+$ wants to be considered instead.

\end{remark}

\subsection{PDF and CDF}

We now show that the \ac{PDF} and \ac{CDF} of the \ac{IFTR} distribution can also be obtained in closed-form, provided that any of the parameters $m_1$ or $m_2$ is restricted to take a positive integer value. We note that the general case of arbitrary real $m_1$ and $m_2$ can always be numerically computed by an inverse Laplace transform over the MGF.

We derive closed-form expressions for the \ac{PDF} and \ac{CDF} of the \ac{SNR} (or, equivalently, the power envelope) for the \ac{IFTR} fading model,  which will be demonstrated in the next lemma. 

\begin{lemma}
When $m_1\in\mathbb{Z}^+$, the \ac{PDF} and \ac{CDF} of the \ac{SNR} $\gamma$ in a \ac{FTR} fading channel can be expressed in terms of the confluent hypergeometric function $\Phi_2(\cdot)$
defined in \cite[p. 34, (8)]{Srivastava1985}, as given, respectively, in (\ref{eq:36}) and (\ref{eq:37}).

\begin{figure*}[!t]
\normalsize
\begin{equation} \label{eq:36}
\begin{split}
  & f_\gamma(x) = \frac{{1 + K}}
{{\bar \gamma }}m_1^{m_1 } m_2^{m_2 } \left( {m_1  + \frac{K}
{2}\left( {1 + \sqrt {1 - \Delta ^2 } } \right)} \right)_{}^{m_2  - m_1 } \sum\limits_{n = 0}^{m_1  - 1} {\frac{1}
{{n!}}\binom{m_1-1}{n}} \frac{{\Gamma \left( {m_2  + n} \right)}}
{{\Gamma \left( {m_2 } \right)}}\left( {\frac{{K\Delta }}
{2}} \right)^{2n}   \cr 
  & \quad \quad  \times \left[ {m_1 \frac{K}
{2}\left( {1 - \sqrt {1 - \Delta ^2 } } \right) + m_2 \frac{K}
{2}\left( {1 + \sqrt {1 - \Delta ^2 } } \right) + m_1 m_2 } \right]^{ - m_2  - n}   \cr 
  & \quad \quad  \times \;\Phi _2^{(3)} \left( {n + 1 - m_1 ,m_1  - m_2 ,m_2  + n;1; - \frac{{1 + K}}
{{\bar \gamma }}x, - \frac{{m_1 (1 + K)}}
{{\left(m_1  + \frac{K}
{2}\left( {1 + \sqrt {1 - \Delta ^2 } } \right)\right)\bar \gamma }}x,} \right.  \cr 
  & \quad \quad \quad \quad \quad \quad \quad \quad \quad \quad \quad \quad \quad \quad \quad \quad \left. { - \frac{{m_1 m_2 (1 + K)}}
{{\left(m_1 \frac{K}
{2}\left( {1 - \sqrt {1 - \Delta ^2 } } \right) + m_2  \frac{K}
{2}\left( {1 + \sqrt {1 - \Delta ^2 } } \right)+m_1 m_2 \right)\bar \gamma }}x} \right). \cr
\end{split}
\end{equation}
\hrulefill
\vspace*{4pt}
\end{figure*}

\begin{figure*}[!t]
\normalsize
\begin{equation} \label{eq:37}
\begin{split}
  & F_\gamma  (x) = \frac{{1 + K}}
{{\bar \gamma }}m_1^{m_1 } m_2^{m_2 } \left( {m_1  + \frac{K}
{2}\left( {1 + \sqrt {1 - \Delta ^2 } } \right)} \right)_{}^{m_2  - m_1 } \sum\limits_{n = 0}^{m_1  - 1} {\frac{1}
{{n!}}\binom{m_1-1}{n}} \frac{{\Gamma \left( {m_2  + n} \right)}}
{{\Gamma \left( {m_2 } \right)}}\left( {\frac{{K\Delta }}
{2}} \right)^{2n}   \cr 
  & \quad \quad  \times \left[ {m_1 \frac{K}
{2}\left( {1 - \sqrt {1 - \Delta ^2 } } \right) + m_2 \frac{K}
{2}\left( {1 + \sqrt {1 - \Delta ^2 } } \right) + m_1 m_2 } \right]^{ - m_2  - n}   \cr 
  & \quad \quad  \times \;x\Phi _2^{(3)} \left( {n + 1 - m_1 ,m_1  - m_2 ,m_2  + n;2; - \frac{{1 + K}}
{{\bar \gamma }}x, - \frac{{m_1 (1 + K)}}
{{\left(m_1  + \frac{K}
{2}\left( {1 + \sqrt {1 - \Delta ^2 } } \right)\right)\bar \gamma }}x,} \right.  \cr 
  & \quad \quad \quad \quad \quad \quad \quad \quad \quad \quad \quad \quad \quad \quad \quad \quad \left. { - \frac{{m_1 m_2 (1 + K)}}
{{\left(m_1 \frac{K}
{2}\left( {1 - \sqrt {1 - \Delta ^2 } } \right) + m_2  \frac{K}
{2}\left( {1 + \sqrt {1 - \Delta ^2 } } \right)+m_1 m_2 \right)\bar \gamma }}x} \right). \cr
\end{split}
\end{equation}
\hrulefill
\vspace*{4pt}
\end{figure*}

\end{lemma}

\begin{IEEEproof}
See Appendix \ref{App3}.
\end{IEEEproof}

Note that despite requiring the evaluation of a confluent hypergeometric function, the \ac{PDF} and \ac{CDF} of the \ac{IFTR} fading model can be expressed in terms of a well-known function in communication theory. In fact, the $\Phi_2(\cdot)$ function also makes appearances in the \ac{CDF} of common fading models such as Rician-shadowed or $\kappa$-$\mu$ shadowed \cite{Paris2010,Paris2014}. Moreover, this function can be efficiently evaluated using a numerical inverse Laplace transform \cite{Martos16}. Thus, the evaluation of the \ac{IFTR} distribution functions does not pose any additional challenge compared to other state-of-the-art fading models, including the FTR fading model.

\begin{figure}[t]
	\centering
	\includegraphics[width=\linewidth]{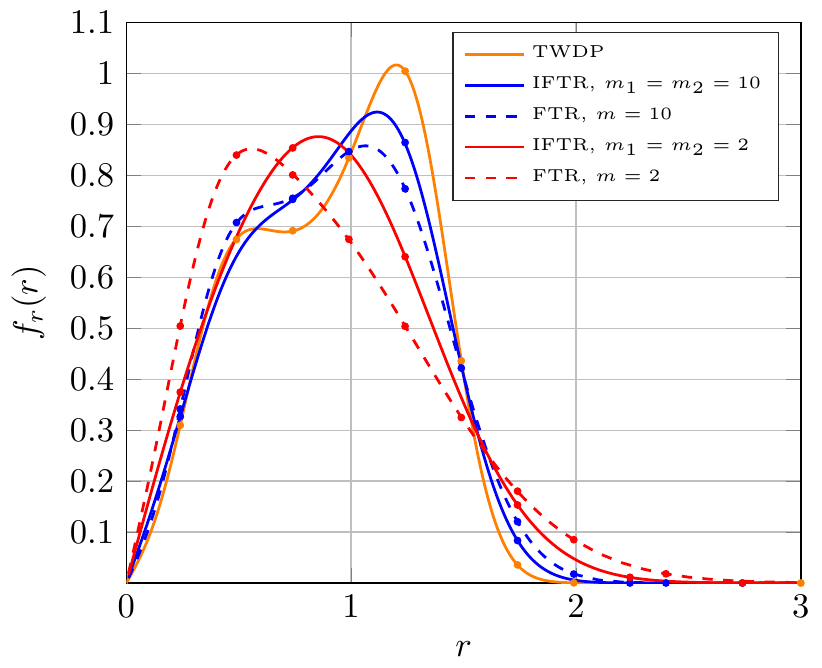}
	\caption{\ac{PDF} of the received signal envelope under \ac{FTR} and \ac{IFTR} fading for $K=15$, $\Delta=0.9$, $\Omega=1$ and different values of $m=m_1=m_2$. Markers correspond to simulation results.}
	\label{fig1}
\end{figure}

Also note that the \ac{PDF} and \ac{CDF} of the received signal envelope, $r=\left|V_r\right|$, can be easily derived from (\ref{eq:36}) and (\ref{eq:37}) by a simple change of variables. Specifically, we can write $f_r(r)=2r f_{\gamma}(r^2)$ and $F_r(r)=F_{\gamma}(r^2)$, with $\bar\gamma$ in (\ref{eq:36}) and (\ref{eq:37}) replaced by $\Omega=\mathbb E\{r^2\}$. In order to illustrate the influence of the independent fluctuation of the specular components on the fading statistics, the \ac{PDF} of the received signal envelope $r$ of the \ac{IFTR} and \ac{FTR} models are compared for the same level of fluctuation, $m=m_1=m_2$. Results for $m=2$ and $m=10$ are shown in Fig. \ref{fig1}. As expected, differences between both \acp{PDF} are larger for low values of $m$ and $(m_1,m_2)$, which correspond to larger fluctuations in the specular components, i.e., larger fading severity. As $m, m_1, m_2 \rightarrow \infty$, the fluctuation decreases and both models tend to the \ac{TWDP} model.  

\begin{figure}[t]
	\centering
	\includegraphics[width=\linewidth]{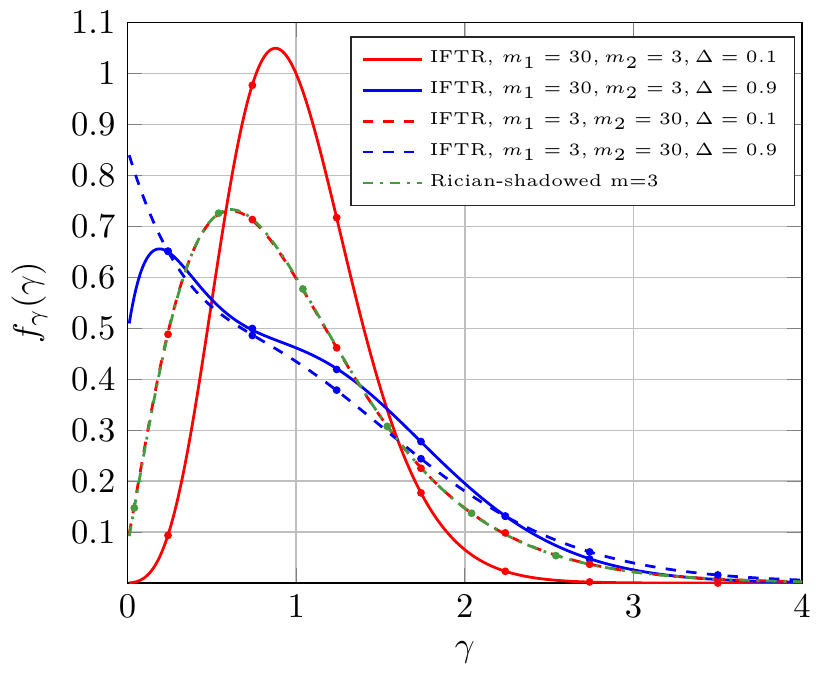}
	\caption{\ac{PDF} of the \ac{SNR} under Rician-shadowed and \ac{IFTR} fading with $\bar \gamma=1, K=15$ and different values of $m_1$, $m_2$ and $\Delta$. Markers correspond to simulation results.}
	\label{fig2}
\end{figure}

The asymmetric fluctuation of the specular components gives the \ac{IFTR} model a remarkable flexibility to change the shape of the \ac{PDF}. This is shown in Fig. \ref{fig2}, where the \ac{PDF} of the \ac{SNR} is depicted for different values of $m_1$ and $m_2$ when $\Delta=0.9$ and $\Delta=0.1$. In all cases $K=15$ and $\bar \gamma=1$. For reference, it is also represented the \ac{PDF} of the Rician-shadowed model with $m=3$ \cite{Abdi2003}. 
It can be seen that, when $\Delta=0.1$, one of the specular components dominates ($V_1 \gg V_2$) and the \ac{PDF} of the \ac{IFTR} distribution tends to that of the Rician-shadowed model with $m=m_1$. When $\Delta=0.9$, both specular components have similar amplitudes and the differences with the Rician-shadowed distribution are noticeable. 

\begin{figure}[t]
	\centering
	\includegraphics[width=\linewidth]{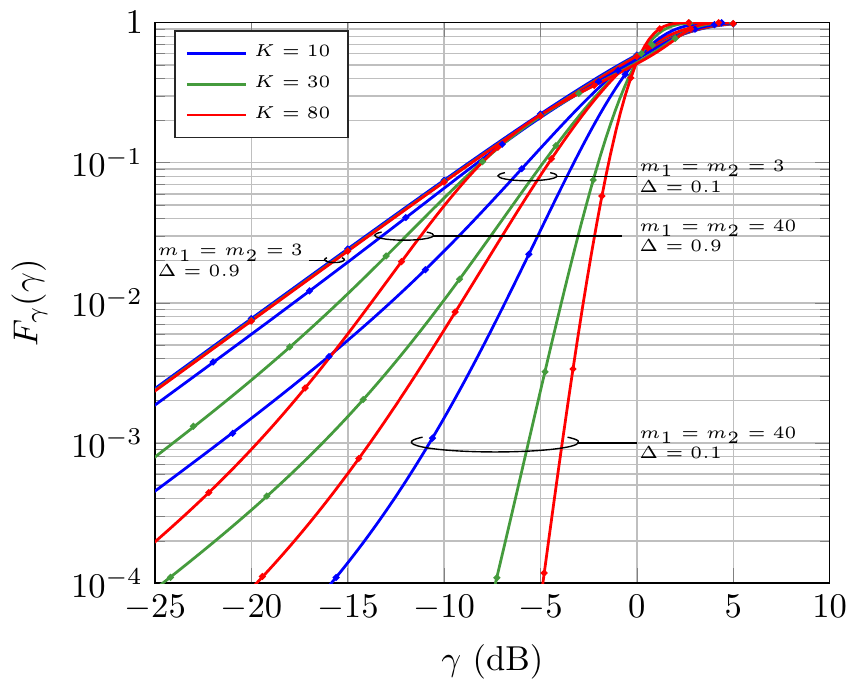}
	\caption{\ac{CDF} of the \ac{SNR} under \ac{IFTR} fading for $\bar \gamma=1$ and different values of the model parameters. Markers correspond simulation results.}
	\label{fig3}
\end{figure}

 
The independent fluctuation of the specular components results in an increased ability of the \ac{CDF} to modify its log-log convexity and concavity, even when $m_1=m_2$. This can be observed in Fig. \ref{fig3}, where the \ac{CDF} of the \ac{SNR} for different values of the \ac{IFTR} parameters are shown. The probability of severe fading is higher for $\Delta=0.9$ because the amplitude of the specular components is similar, and therefore destructive multipath combination occurs more likely. The probability of the specular components to cancel each other is also higher for low values of $m_1$ and $m_2$, i.e., when the fluctuations are larger. In these circumstances, the influence of $K$ is small, since the modulus of the sum $\sqrt {\zeta _1 } V_1  \exp \left( {j\phi _1 } \right) +  \sqrt {\zeta _2 } V_2 \exp \left( {j\phi _2 } \right)$ in (\ref{eq:01}) will be generally small with respect to the power of the diffuse term, irrespectively of the magnitude of $V_1^2+V_2^2$. Conversely, deep-fading probability reduces as $\Delta$ decreases, because one of the specular components becomes much larger than the other.


\section{Empirical validation}
\label{fit}

This section illustrates the capability of the \ac{IFTR} fading to model the small-scale fading in quite diverse scenarios. Three types of outdoor communication links are considered: the \ac{LOS} \ac{mmWave} channels given in \cite{Samimi2016} and \cite{Sun17}, the \ac{LMS} channels in \cite{Abdi2003} and the \ac{UAC} channels measured in \cite{Sanchez14,Canete16}. It will be shown that the \ac{IFTR} model provides a better fit to the experimental datasets than the models previously used in each of these scenarios, and also than the \ac{FTR} model. 

The goodness of fit between the analytical \ac{CDF} of the considered model, $F_a(x)$, and the empirical one estimated from measurements, $F_e(x)$, is quantified by using the following modified version of the \ac{KS} statistic:

\begin{equation}   
\epsilon=\max_{x}\left|\log_{10}F_e(x)-\log_{10}F_a(x)\right|.
\label{eq.fit.1}
\end{equation}

The logarithm in (\ref{eq.fit.1}) is used to magnify the errors between the \acp{CDF} in the region close to zero. The rationale to overweight these values is that some of the key performance metrics of communication systems, e.g., the \ac{BER} and the outage probability, are determined by the probability of deep fading events. Hence, improving the goodness of the fitting in this region is more important.

The number of parameters of the \ac{IFTR} distribution to be optimized depends on the considered experimental dataset. In the \ac{mmWave} and \ac{UAC} channels, the empirical \ac{CDF} of $x=r^2/\mathbb E\{r^2\}$ is given \cite{Samimi2016,Sun17,Sanchez14, Canete16}, while the one of $x=r^2$ is reported in \cite{Abdi2003}. Hence, in addition to $m_1, m_2, K$ and $\Delta$, the optimum value of $\Omega=\mathbb E\{r^2\}$ has to be determined also in the latter. Similar considerations apply to the remaining models included in the analysis. 
                 
\subsection{mmWave channels}

Two outdoor \ac{LOS} radio channels are considered in this subsection: a cross-polarized channel measured at 28 GHz, from now on referred to as Ch. 1, and a co-polarized one measured at 73 GHz, referred to as Ch. 2. The empirical \ac{CDF}s of their small-scale fading amplitudes are given in \cite[Fig. 6]{Samimi2016} and \cite[Fig. 3]{Sun17}, respectively. The capability of the \ac{IFTR} fading to model these channels is compared to that of the Rice distribution, which has been previously used to model \ac{LOS} channels in these bands. Moreover, the \ac{TWDP} fading is also considered for completeness, since the analysis in \cite{Zochmann19} supports that it is favored over the Rice one. As both the \ac{IFTR} and the \ac{FTR} models generalize the \ac{TWDP} one, the comparison is useful to assess the potential gain given by the fluctuations in the specular components in terms of fitting quality. 

Table \ref{Table.1} shows the fitting error and the optimum model parameters for the aforementioned channels. For the case of Ch. 1 the \ac{IFTR} fading model yields a lower error than the \ac{FTR} one, which already improves the results achieved by the \ac{TWDP} fading model. We see that the Rice model provides the largest error. In Ch. 2 the improvement of the \ac{IFTR} model with respect to the remaining models is even higher, and is actually quite significant. Interestingly, the \ac{TWDP} and the \ac{FTR} models achieve the same error as with the Rice model only in their limiting cases, i.e., $\Delta=0$ in the \ac{TWDP} model and ($\Delta=0,m \rightarrow \infty$) in the \ac{FTR} one. This might erroneously lead to conclude that the second specular component and the \ac{LOS} fluctuation are superfluous to model this channel. However, the results achieved by the \ac{IFTR} fading model suggest otherwise, indicating that the consideration of \textit{both} components is beneficial as long as their fluctuation level can be independently controlled. Hence, the large value of $m_1$ suggests that the specular component with larger amplitude experiences almost no fluctuation, i.e., $\zeta_1$ is almost constant, while the lower amplitude component $\sqrt{\zeta_2}V_2$ fluctuates considerably. 

\begin{table}[t]
\renewcommand{\arraystretch}{1.3}
\caption{Fitting results for the mmWave channels}
\label{Table.1}
\centering
\begin{tabular}{c@{\hspace{0.2cm}}ccccc}
& & \multicolumn{4}{c}{\textbf{Model}}\\
\cline{3-6}
\textbf{Channel} & \textbf{Param.} & IFTR & FTR & TWDP & Rice\\
\hline
\multirow{4}*{Ch. 1} & $\epsilon$ &  0.2203 & 0.2246 & 0.2267 & 0.3298 \\
& $K$ & 476.1454& 80.3916& 23.1347& 3.5820\\
& $\Delta$ & 0.8463 & 0.5873 & 0.8619 & -\\
& $\left(m_1,m_2\right)/m$ & (9, 50.5) & 2 & - & -\\[0.2cm]
\hline
\multirow{4}*{Ch. 2} &$\epsilon$ & 0.2068 & 0.3228 & 0.3228 & 0.3228 \\
&$K$ & 154.3797 & 45.4773 & 45.4773 & 45.4773 \\
&$\Delta$ & 0.2170 & 0.0000 & 0.0000 & -\\
&$\left(m_1,m_2\right)/m$ & (60, 3.6) & $\infty$ & - & -\\
\hline
\end{tabular}
\end{table}


\begin{figure}[t]
	\centering
	\includegraphics[width=.98\columnwidth]{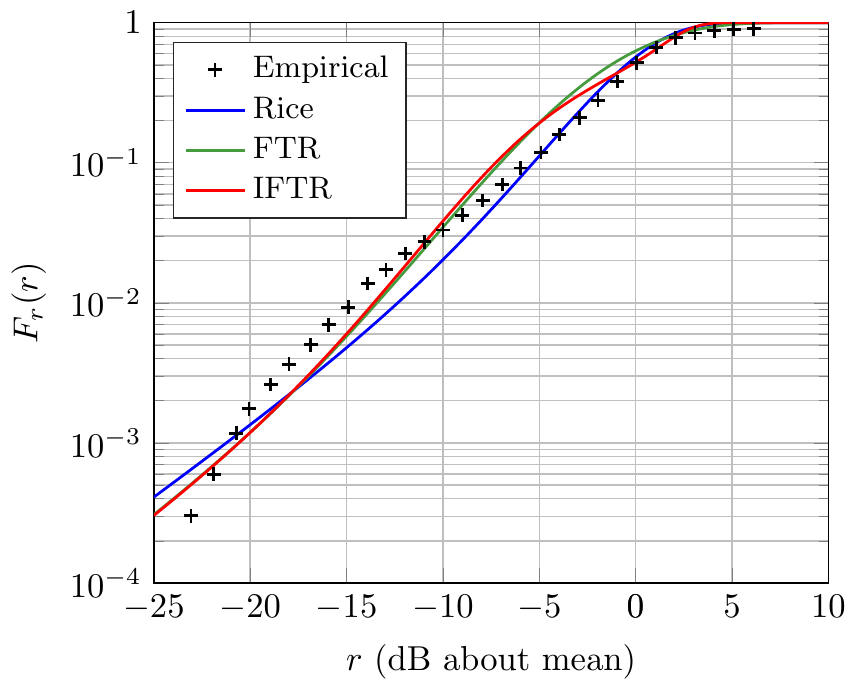}
	\caption{Empirical and theoretical \acp{CDF} of the received signal amplitude for \ac{mmWave} Ch. 1 \cite[Fig. 8]{Romero2017}.}
	\label{Fig.CDF_Samimi16}
\end{figure}

The enhanced modeling capacity of the \ac{IFTR} model can be graphically interpreted in terms of its flexibility to modify the concavity and convexity of the \ac{CDF} (in a log-log scale). We illustrate this fact by reproducing the empirical and theoretical \acp{CDF} of the Rice and \ac{FTR} models for Ch. 1 given in \cite[Fig. 8]{Romero2017}, along with the \ac{CDF} of the \ac{IFTR} model. Results are shown in Fig. \ref{Fig.CDF_Samimi16}, where it can be seen that the \ac{CDF} of the \ac{IFTR} model is convex for $0<r<5$ dB, concave for $r$ around $-5$ dB, and convex again for $r<-17.5$ dB. 

\subsection{LMS channels}

The small-scale fading in \ac{LMS} channels has been widely modeled using the Rician-shadowed distribution, where the \ac{LOS} component is assumed to undergo Nakagami-$m$ fading. The IFTR model can be seen as a generalization of the Rician-shadowed one. The capability of the \ac{IFTR} fading to outperform the Rician-shadowed fading model is assessed by using the channels in \cite[Fig. 1]{Abdi2003}, where the empirical \ac{CDF}s of two channels at 870 MHz are provided: one experiencing light shadowing, and another one affected by heavy shadowing caused by a denser tree cover.

Table \ref{Table.2} shows the fitting results for the \ac{IFTR}, \ac{FTR} and Rician-shadowed models. For the sake of fairness, the parameters of the latter (denoted using the subscript RS) have been computed to minimize the error in (\ref{eq.fit.1}), as the ones provided in \cite{Abdi2003} were obtained with a different metric. It can be seen that the \ac{IFTR} model yields better fitting results than the other models for both channels. In the light shadowing channel, the high value of $m_{\rm RS}$ suggests that the \ac{LOS} term is essentially constant. Moreover, the \ac{FTR} model achieves the error of the Rician-shadowed one only in its limiting case, $\Delta=0$ and $m \rightarrow \infty$. These facts might again lead to conclude that a single specular component with constant amplitude suffices to adequately model this channel. In fact, the Rice distribution with $K=3.1075$ yields the same error as these two models (although not explicitly shown in Table \ref{Table.2}). However, the values ($m_1=14, m_2=0.8)$ of the \ac{IFTR} model indicate that two specular components, with considerable fluctuation in the one with lower amplitude, may be more adequate for a proper channel modeling. 

\begin{figure}[t]
	\centering
	\includegraphics[width=1\columnwidth]{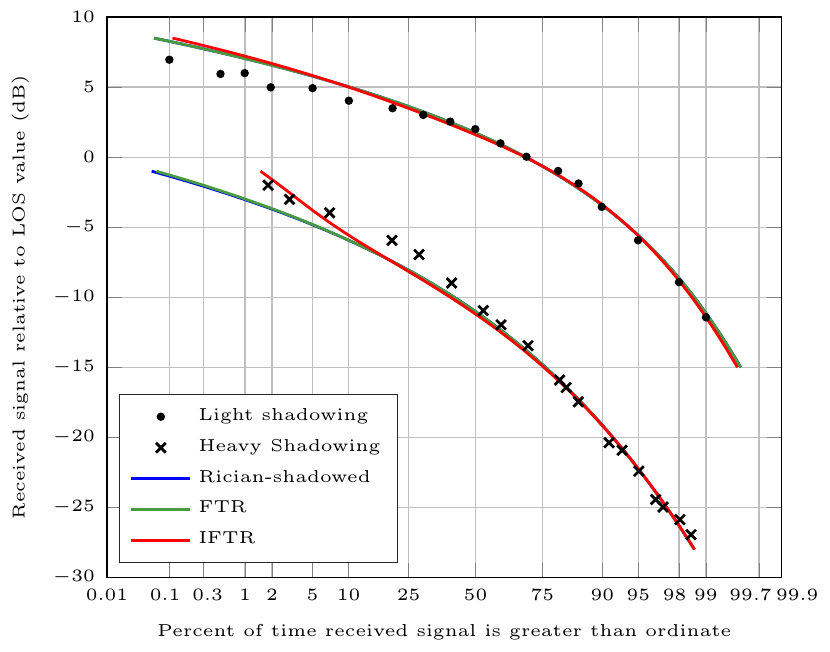}
	\caption{Empirical and theoretical \acp{CCDF} of the received signal amplitude for the light and heavy shadowing \ac{LMS} channels given in \cite[Fig. 1]{Abdi2003}.}
	\label{Fig.CDF_Abdi03}
\end{figure}

The improved capability of the \ac{IFTR} distribution to model the two considered channels can be graphically observed in Fig. \ref{Fig.CDF_Abdi03}, where the empirical and theoretical \acp{CCDF} of the received signal envelope are shown. For coherence, axes have been scaled as in \cite[Fig. 1]{Abdi2003}. Hence, errors have to be measured as the horizontal distance between the empirical and theoretical (logarithm of the) \ac{CCDF} for a given value of the received signal level. In the heavily-shadowed channel, where all models fit the \ac{CCDF} almost equally well in the low signal level region, the flexibility of the \ac{IFTR} model is clearly noticeable in the high signal level region, where its \ac{CCDF} changes from (log-log) convex to concave as the received signal level decreases. In the light-shadowing channel, the lower error of the \ac{IFTR} model is due to its ability to achieve a better trade-off than the remaining models between the fitting in the low signal level region, where errors are overweighted, and in the high signal level zone, where they have been considered less relevant. 

\begin{table}[t]
\renewcommand{\arraystretch}{1.3}
\caption{Fitting results for the \ac{LMS} channels}
\label{Table.2}
\centering
\begin{tabular}{c@{\hspace{0.2cm}}cccc}
& & \multicolumn{3}{c}{\textbf{Model}}\\
\cline{3-5}
\textbf{Channel} & \textbf{Parameter} & IFTR & FTR & \parbox[c]{1.2cm}{\vspace{0.1cm}\centering{Rician shadowed}\vspace{0.1cm}}\\
\hline
\hspace{-0.7cm}\multirow{5}*{\parbox[c]{0.2cm}{\centering{Light shad.}}} & $\epsilon$ &  0.0442 & 0.0507 & 0.0507 \\
& $K$ & 479.8224& 4.0542& -\\
& $\Delta / b_{0_{\rm RS}}$ & 0.8290 & 0 & 0.2025\\
& $\Omega/\Omega_{\rm RS}$ & 1.6692 & 1.6896 & 1.2847\\
& $\left(m_1,m_2\right)/m/m_{\rm RS}$ & (14, 0.8) & $\infty$ & 100\\[0.2cm]
\hline
\hspace{-0.7cm}\multirow{5}*{\parbox[c]{0.2cm}{\centering{Heavy shad.}}} &$\epsilon$ & 0.0655 & 0.0672 & 0.0673 \\
& $K$ & 2.7457& 0.7484& -\\
& $\Delta / b_{0_{\rm RS}}$ & 0.9997 & 0.6040 & 0.0429\\
& $\Omega/\Omega_{\rm RS}$ & 0.1289 & 0.1121 & 0.0258 \\
& $\left(m_1,m_2\right)/m/m_{\rm RS}$ & (2, 0.1) & 2 & 100\\
\hline
\end{tabular}
\end{table}

\subsection{UAC channels}
\ac{UAC} channels in shallow waters exhibit multipath propagation and fading. The former is due to the reflections in the seabed, the surface of the sea and other objects, while the latter is caused by the surface waves (which change the reflection angles), internal waves and the motion of the communication ends, among other factors \cite{Walree13}. 

The $\kappa$-$\mu$ shadowed distribution, which generalizes the Rice and Rician-shadowed ones, has been proposed to model these effects \cite{Canete16}. However, $\mu\in\mathbb{R}$ values are employed to this end. This considerably hinders the physical interpretation of the model (which requires $\mu\in\mathbb{Z}^+$ according to the physical model definition in \cite{Paris2014}), and the generation of random variables for simulation purposes \cite{Yacoub07,Paris2014}. Hence, it is of practical interest to assess its fitting capabilities also when constraining $\mu \in \mathbb{Z}^+$. It must be recalled that, for practical purposes, the \ac{IFTR} model employed in this section is also constrained to have $m_1 \in \mathbb{Z}^+$, to allow for the analytical calculation of the \ac{PDF}, despite random variables following the \ac{IFTR} distribution can be easily generated according to (\ref{eq:01}) for non-negative real values of $m_1$. The same applies to the \ac{FTR} model, where $m \in \mathbb{Z}^+$ is also considered.  

The ability of the \ac{IFTR} model to fit \ac{UAC} channels is assessed by using the measurements in \cite{Sanchez14,Canete16}. They provide a characterization of 10 channels corresponding to links with lengths 50, 100 and 200 m, measured in Mediterranean shallow waters with a sandy seabed and depths from 14 to 30 m, approximately. The transmitter and receiver where placed at depths 3, 6 and 9 m. Sounding signals of 32, 64 and 128 kHz were transmitted. For coherence with \cite{Sanchez14,Canete16}, channels are denoted using the nomenclature [A$|$B$|$C]D-F, where the A$|$B$|$C denotes the link length (50$|$100$|$200 m, respectively), D indicates the transmitter-receiver depth in meters and F the frequency of the transmitted signal in kHz. 

The \ac{IFTR} model fits these 10 channels better than the $\kappa-\mu$ shadowed model with $\mu \in \mathbb{Z}^+$, except for the case of C9-32, where it yields the same error. For conciseness, only the results obtained in one channel of each length (50, 100 and 200 m) are shown in Table \ref{Table.3}, where the values corresponding to the \ac{FTR} model are also shown. As in the previous scenarios, the flexibility of the \ac{IFTR} fading model to control the fluctuations in each specular component independently is also exploited in this context. In channels A6-32 and B6-64 the specular component with lower amplitude is almost constant, and the component with larger amplitude exhibits considerable fluctuations. In channel C3-64, the fluctuation of the specular component with larger amplitude is much lower than the one with lower amplitude, but still considerable. 

\begin{figure}[htb]
	\centering
	\includegraphics[width=1\columnwidth]{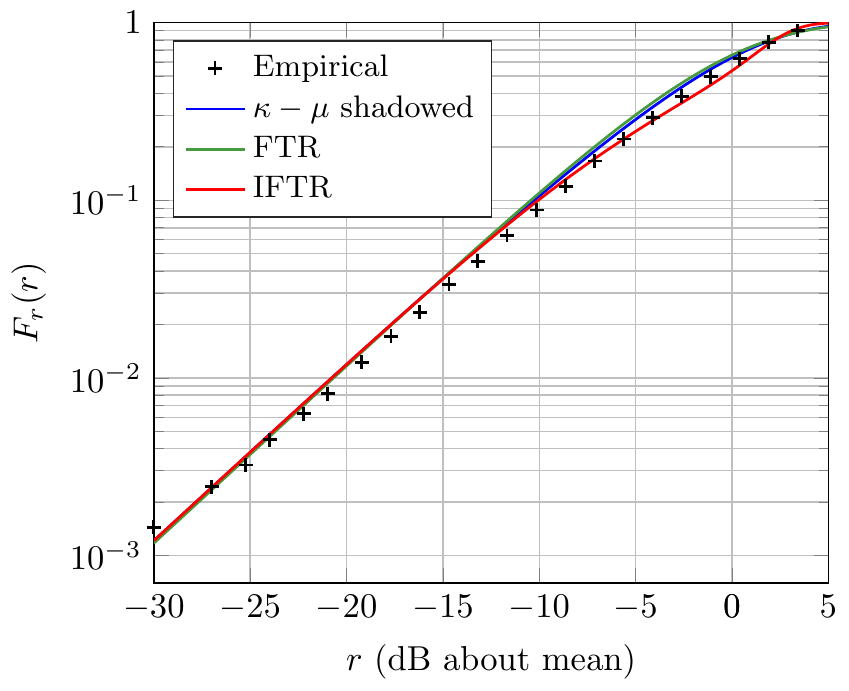}
	\caption{Empirical and theoretical \acp{CDF} of the received signal amplitude for the B6-64 \ac{UAC} channel given in \cite{Sanchez14, Canete16}.}
	\label{Fig.CDF_UAC_B6_64}
\end{figure}

As illustrated in the previous scenarios, the independent fluctuation of the specular components results in an increased ability of the \ac{CDF} to change from concave to convex, and vice-versa. This can be observed in Fig. \ref{Fig.CDF_UAC_B6_64}, where the empirical and theoretical \acp{CDF} corresponding to channel B6-64 are shown. It can be seen that the \ac{CDF} of the \ac{IFTR} distribution is (log-log) concave around $r=5$ dB, convex around $r=0$ dB, and concave again for $r<-5$ dB.

\begin{table}[htb]
\renewcommand{\arraystretch}{1.3}
\caption{Fitting results for the \ac{UAC} channels}
\label{Table.3}
\centering
\begin{tabular}{c@{\hspace{0.2cm}}cc@{\extracolsep{0.3cm}}c@{\extracolsep{0cm}}c}
& & \multicolumn{3}{c}{\textbf{Model}}\\
\cline{3-5}
\textbf{Channel} & \textbf{Parameter} & IFTR & FTR & \parbox[c]{1.2cm}{\vspace{0.1cm}\centering{$\kappa-\mu$ shadowed}\vspace{0.1cm}}\\
\hline
\multirow{5}*{A6-32} & $\epsilon$ &  0.0569 & 0.0622 & 0.0682 \\
& $K/\kappa$ & 466.2619& 4.8472& 1.9494\\
& $\Delta/\mu$ & 0.8720 & 0.9510 & 1\\
& $\left(m_1,m_2\right)/m/m_{\kappa-\mu}$ & (2, 60) & 24 & 1.3088\\[0.2cm]
\hline
\multirow{5}*{B6-64} & $\epsilon$ &  0.0746 & 0.0886  & 0.0758 \\
& $K/\kappa$ & 508.9355 & 499942.1713 & 51.3649\\
& $\Delta/\mu$ & 0.9598 & 0.5256 & 1\\
& $\left(m_1,m_2\right)/m/m_{\kappa-\mu}$ & (4, 60) & 1 & 0.9360\\[0.2cm]
\hline
\multirow{5}*{C3-64} & $\epsilon$ &  0.1880 & 0.1969  & 0.1979 \\
& $K/\kappa$ & 501.1807 & 5.4941 & 6.3239\\
& $\Delta/\mu$ & 0.5662 & 0 & 1\\
& $\left(m_1,m_2\right)/m/m_{\kappa-\mu}$ &(7, 0.75) & $\infty$ & 24.2813\\[0.2cm]
\hline
\end{tabular}
\end{table}

\section{Performance analysis of wireless communications systems}
\label{performance}

After providing a thorough empirical validation of the IFTR fading model in a number of scenarios, it is time to illustrate how the exact closed-form expressions of the statistics of the \ac{SNR} can be used for performance analysis. For exemplary purposes, we calculate the \ac{BER} for a family of coherent modulations, and the outage probability.

\subsection{Average BER}

The \ac{CEP}, i.e., the error rate under \ac{AWGN}, for many wireless communication systems with coherent detection is determined by \cite{Lopez2010}
\begin{equation}
\label{eq:43}
P_E \left( \gamma \right) = \sum\limits_{r = 1}^R {\alpha _r Q\left( {\sqrt {\beta _r \gamma} } \right)}, 
\end{equation}
where $Q(\cdot)$ is the Gauss $Q$-function and $\left\{ {\alpha _r ,\beta _r } \right\}_{r = 1}^R $ are modulation-dependent constants.

The average error rates for the \ac{CEP} given in (\ref{eq:43}) can be calculated in terms of the \ac{CDF} of the \ac{SNR} as \cite{Romero2017}
\begin{equation} \label{eq:42}
	\overline{P_e}=\sum\limits_{r = 1}^R \alpha _r \int^{\infty}_{0}   { \sqrt {\frac{{\beta _r }}
{{8\pi x}}} } e^{ - \frac{{\beta _r x}}
{2}} F_\gamma (x) dx.
\end{equation}
Introducing (\ref{eq:37}) into (\ref{eq:42}), and with the help of \cite[p. 286, (43)]{Srivastava1985}, a compact exact expression of the average \ac{BER} can be found, as given in (\ref{eq:45}), in terms of the Lauricella function $F_D(\cdot)$ defined in \cite[p. 33, (4)]{Srivastava1985}.
\begin{figure*}[!t]
\normalsize
\begin{equation}
\label{eq:45}
\begin{split}
  & \overline{P_e}  = \frac{{1 + K}}
{{\bar \gamma }}m_1^{m_1 } m_2^{m_2 } \left( {m_1  + \frac{K}
{2}\left( {1 + \sqrt {1 - \Delta ^2 } } \right)} \right)_{}^{m_2  - m_1 } \sum\limits_{n = 0}^{m_1  - 1} {\frac{1}
{{n!}}\binom{m_1-1}{n}} \frac{{\Gamma \left( {m_2  + n} \right)}}
{{\Gamma \left( {m_2 } \right)}}\left( {\frac{{K\Delta }}
{2}} \right)^{2n}   \cr 
  & \quad \quad  \times \left[ {m_1 \frac{K}
{2}\left( {1 - \sqrt {1 - \Delta ^2 } } \right) + m_2 \frac{K}
{2}\left( {1 + \sqrt {1 - \Delta ^2 } } \right) + m_1 m_2 } \right]^{ - m_2  - n} \sum\limits_{r = 1}^R {\alpha _r \frac{1}
{{2\beta _r }}}   \cr 
  & \quad \quad  \times \;F_D^{(3)} \left( {\frac{3}
{2},n + 1 - m_1 ,m_1  - m_2 ,m_2  + n;2;  - \frac{{2(1 + K)}}
{{\beta _r \bar \gamma }}, - \frac{{2m_1 (1 + K)}}
{{\beta _r \left( {m_1  + \frac{K}
{2}\left( {1 + \sqrt {1 - \Delta ^2 } } \right)} \right)\bar \gamma }},} \right.  \cr 
  & \quad \quad \quad \quad \quad \quad \quad \quad \quad \quad \quad \quad \quad \quad \quad \quad \quad \quad \left. { - \frac{{2m_1 m_2 (1 + K)}}
{{\beta _r \left( {m_1 \frac{K}
{2}\left( {1 - \sqrt {1 - \Delta ^2 } } \right) + m_2  \frac{K}
{2}\left( {1 + \sqrt {1 - \Delta ^2 } } \right)+m_1 m_2 } \right)\bar \gamma }}} \right). \cr
\end{split}
\end{equation}
\hrulefill
\vspace*{4pt}
\end{figure*}

Although the derived \ac{BER} expression can be easily computed using the Euler form of the $F_D$ function, it does not provide insight about the impact of the different system parameters on performance. We now present an asymptotic, yet accurate, simple expression for the error rate in the high \ac{SNR} regime. First, note that the equality in (\ref{eq:46}) holds for the asymptotic behavior of the \ac{MGF}, where we write a function $a(x)$  as $o(x)$ if $\lim_{x\rightarrow\infty} a(x)/x = 0$. Thus, performing a similar approach to that in \cite[Propositions 1 and 3]{Wang03}, and after some algebraic manipulation, the asymptotic \ac{BER} expression given in (\ref{eq:47}) is obtained.
\begin{figure*}[!t]
\normalsize
\begin{equation}
\label{eq:46}
\begin{split}
  &  \left| {M_\gamma  \left( s \right)} \right| = \frac{{1 + K}}
{{\bar \gamma \left| s \right|}}\frac{{m_1^{m_1 } }}
{{\left[ {m_1  + \frac{K}
{2}\left( {1 + \sqrt {1 - \Delta ^2 } } \right)} \right]^{m_1 } }}\frac{{m_2^{m_2 } }}
{{\left[ {m_2  + \frac{K}
{2}\left( {1 - \sqrt {1 - \Delta ^2 } } \right)} \right]^{m_2 } }}  \cr 
  & \quad \quad \times \,_2 F_1 \left( {m_1 ,m_2 ;1;\frac{{K^2 \Delta ^2 }}
{{\left[ {2m_1  + K\left( {1 + \sqrt {1 - \Delta ^2 } } \right)} \right]\left[ {2m_2  + K\left( {1 - \sqrt {1 - \Delta ^2 } } \right)} \right]}}} \right) + o\left( {\left| s \right|^{ - 1} } \right). \cr
\end{split}
\end{equation}
\hrulefill
\vspace*{4pt}
\end{figure*}

\begin{figure*}[!t]
\normalsize
\begin{equation}
\label{eq:47}
\begin{split}
 & \overline{P_e}  \approx \frac{{1 + K}}
{{2\bar \gamma }}\frac{{m_1^{m_1 } }}
{{\left[ {m_1  + \frac{K}
{2}\left( {1 + \sqrt {1 - \Delta ^2 } } \right)} \right]^{m_1 } }}\frac{{m_2^{m_2 } }}
{{\left[ {m_2  + \frac{K}
{2}\left( {1 - \sqrt {1 - \Delta ^2 } } \right)} \right]^{m_2 } }}\left( {\sum\limits_{r = 1}^R {\frac{{\alpha _r }}
{{\beta _r }}} } \right)  \cr 
  & \quad \quad  \times \,_2 F_1 \left( {m_1 ,m_2 ;1;\frac{{K^2 \Delta ^2 }}
{{\left[ {2m_1  + K\left( {1 + \sqrt {1 - \Delta ^2 } } \right)} \right]\left[ {2m_2  + K\left( {1 - \sqrt {1 - \Delta ^2 } } \right)} \right]}}} \right),\quad \bar \gamma  \gg 1. \cr
\end{split}
\end{equation}
\hrulefill
\vspace*{4pt}
\end{figure*}

\subsection{Outage probability}

The instantaneous channel capacity per unit bandwidth is well-known to be given by
\begin{equation}
\label{eq:48}
C = \log _2 (1 + \gamma ).
\end{equation}
We define the outage capacity probability, or simply outage probability, as the probability that the instantaneous channel capacity $C$ falls below a predefined threshold  $R_S$ (given in terms of rate per unit bandwidth), i.e.,
\begin{equation}
\label{eq:49}
P_{\rm out}  = P\left( {C < R_S } \right) = P\left( {\log _2 (1 + \gamma ) < R_S } \right).
\end{equation}
Therefore
\begin{equation}
\label{eq:50}
P_{\rm out}  = P\left( {\gamma  < 2^{R_S }  - 1} \right) = F_\gamma  \left( {2^{R_S }  - 1} \right).
\end{equation}
Thus, the outage probability can be directly calculated from (\ref{eq:37}) specialized for $x=2^{R_S } - 1$. This expression is exact and holds for all \ac{SNR} values; however, it offers little insight about the effect of parameters on system performance. Fortunately, we can obtain a simple expression in the high \ac{SNR} regime as follows: From (\ref{eq:37}) and \cite[Proposition 5]{Wang03}, the \ac{CDF} of $\gamma$ can be written as given in (\ref{eq:51}). Therefore, the outage probability can be approximated in the large \ac{SNR} regime by introducing (\ref{eq:51}) into (\ref{eq:50}).
\begin{figure*}[!t]
\normalsize
\begin{equation}
\label{eq:51}
\begin{split}
  & F_\gamma  (x) = \frac{{1 + K}}
{{\bar \gamma }}\frac{{m_1^{m_1 } }}
{{\left[ {m_1  + \frac{K}
{2}\left( {1 + \sqrt {1 - \Delta ^2 } } \right)} \right]^{m_1 } }}\frac{{m_2^{m_2 } }}
{{\left[ {m_2  + \frac{K}
{2}\left( {1 - \sqrt {1 - \Delta ^2 } } \right)} \right]^{m_2 } }}  \cr 
  & \quad \quad \quad  \times \,_2 F_1 \left( {m_1 ,m_2 ;1;\frac{{K^2 \Delta ^2 }}
{{\left[ {2m_1  + K\left( {1 + \sqrt {1 - \Delta ^2 } } \right)} \right]\left[ {2m_2  + K\left( {1 - \sqrt {1 - \Delta ^2 } } \right)} \right]}}} \right)x + o\left( {\bar \gamma ^{-1}}\right). \cr
\end{split}
\end{equation}
\hrulefill
\vspace*{4pt}
\end{figure*}

\section{Numerical results}
\label{numerical}
In this section, numerical results are presented to evaluate the performance metrics of point-to-point wireless systems undergoing IFTR fading. Monte-Carlo simulations are used in all instances to validate the obtained closed-form expressions. 

\begin{figure}[t]
	\centering
	\includegraphics[width=\linewidth]{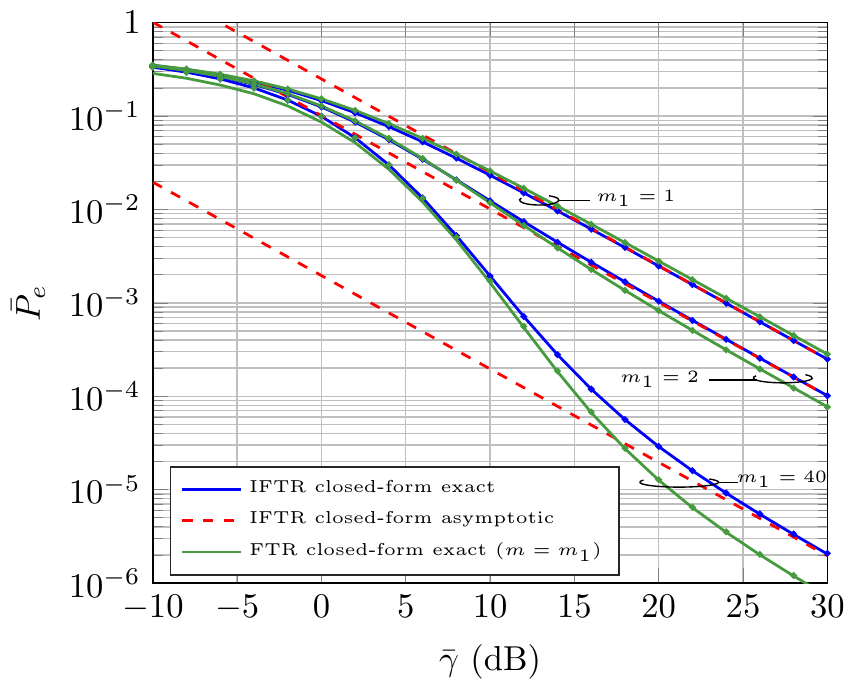}
	\caption{Exact and asymptotic closed-form \ac{BER} vs. average \ac{SNR} for BPSK modulation under \ac{IFTR} and \ac{FTR} fading with $K=15$, $\Delta=0.5$, $m_2=2$ and different values of $m_1=m$. Markers denote results obtained using Monte-Carlo simulations.}
	\label{fig4}
\end{figure}

Figs. \ref{fig4} shows the \ac{BER} of the BPSK modulation versus the average \ac{SNR}. Results obtained with the closed-form exact and asymptotic \ac{BER} expressions given in \eqref{eq:45} and \eqref{eq:47} (with $R=1$, $\alpha_1=1$ and $\beta_1=2$), respectively, are given for $K=15$, $\Delta=0.5$, $m_2=2$ and different values of $m_1$. \ac{BER} values obtained under \ac{FTR} fading with the same $K, \Delta$ and $m=m_1$ are also given as a reference, to better highlight the differences between both models. It can be seen that there is a rather good match between the asymptotic and the exact \ac{BER} in the high \ac{SNR} region, and performance increases when $m_1$ raises due to lower fading severity. Similarly, differences between the \ac{IFTR} and the \ac{FTR} models also increase with $m_1$, being about 3.6 dB for $m_1=m=40$. The worse (higher) \ac{BER} of the \ac{IFTR} fading in this case is due to the large fluctuation experienced by the specular component with amplitude $V_2$, which in the \ac{FTR} fading is almost negligible as $m=40$.

\begin{figure}[t]
	\centering
	\includegraphics[width=\linewidth]{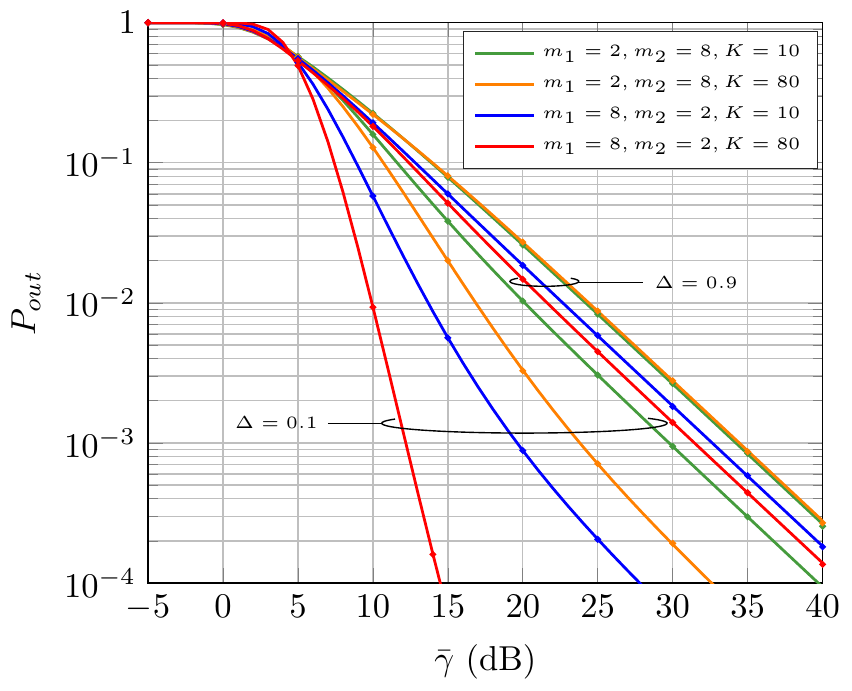}
	\caption{Outage probability vs. average \ac{SNR} for $R_s=2$ and different combinations of $m_,1, m_2, K$ and $\Delta$. Markers denote results obtained using Monte-Carlo simulations.}
	\label{fig5}
\end{figure}

Fig. \ref{fig5} shows the outage probability versus the average \ac{SNR} for $R_s=2$. Results are plotted for $\Delta=0.1$ and $\Delta=0.9$ and various combinations of $m_1, m_2$ and $K$. It can be seen that the outage probability decreases as $\Delta$ becomes lower, since in that case one specular component is much larger than the other one and it is unlikely that they cancel out each other. In these circumstances, performance also improves as $m_1$ increases, because the fluctuation of the largest specular component becomes smaller. It can also be noticed that the influence of $K$ is higher for $\Delta=0.1$, where cancellation of the specular components is unlikely. 

For $\bar \gamma >5$ (dB), the outage probability decreases as $K$ increases, except for $(m_1=2, m_2=8)$ and $\Delta=0.9$, where performance is lower for $K=80$ than for $K=10$. This occurs because having a destructive combination of the specular components becomes more detrimental as $K$ increases, since almost all the received signal power is conveyed by the specular components as $K \to \infty$. When $\Delta<1$, the probability of destructive sum depends on the specific fluctuation of the specular components, and becomes larger as the fluctuation of the larger specular component increases, i.e., $m_1$ decreases. This is reflected by the fact that the outage probability is higher for $(m_1=2, m_2=8)$ than for $(m_1=8, m_2=2)$. The asymmetry in the fluctuation of the specular components becomes immaterial when $\Delta=1$, where $(m_1=2, m_2=8)$ and $(m_1=8, m_2=2)$ yielding the same outage probabilities (not explicitly shown in Fig. \ref{fig5}). 


\section{Conclusion}
\label{conc}
In this work, a new small-scale fading model consisting of two specular components with independent fluctuation plus a diffuse term has been introduced for the first time in the literature. The IFTR fading model innovates over the \ac{FTR} model, whose specular components jointly fluctuate, and generalizes previous models such as the Rice, Rician-Shadowed, \ac{TWDP} and others. The \ac{IFTR} model has a clear physical interpretation, as it models the case where the specular waves follow very different paths and are affected by different scatterers and/or perturbations. The presented results suggest that this situation is quite common in rather diverse scenarios such as \ac{LOS} \ac{mmWave}, \ac{LMS} and \ac{UAC} channels.

Closed-form expressions for the \ac{PDF}, \ac{CDF} and \ac{MGF} of the received SNR, expressed in terms of functions used in other state-of-the-art fading models, have been provided and double-checked. The \ac{IFTR} model is shown to fit channels measured in a number of scenarios better than the \ac{FTR} model and other previously proposed models. The outage probability and the error performance of wireless communication systems over channels with \ac{IFTR} fading have been obtained in closed-form. 

\appendices
\section{Proof of Lemma 1}
\label{App1}

Let us consider the fading channel model given in (\ref{eq:01}) conditioned to a particular realization $\zeta_1=u_1$, $\zeta_2=u_2$ of the random variables modeling the fluctuation of the specular components. In this case, we can write
\begin{equation}
\label{eq:12}
\left. {V_r } \right|_{ u_1, u_2}  = \sqrt u_1 V_1 \exp \left( {j\phi _1 } \right) + \sqrt u_2 V_2 \exp \left( {j\phi _2 } \right) + X + jY.
\end{equation}
This corresponds to the classical TWDP fading model where the amplitudes of the specular components are given by $\sqrt u_1 V_1$ and $\sqrt u_2 V_2$, for which the following ancillary parameters can be defined:
\begin{equation}
\label{eq:13}
K_{u_1 ,u_2 }  = {{u_1 V_1^2  + u_2 V_2^2 } \over {2\sigma ^2 }},
\end{equation}
\begin{equation}
\label{eq:14}
\Delta _{u_1 ,u_2 }  = {{2\sqrt {u_1 u_2 } V_1 V_2 } \over {u_1 V_1^2  + u_2 V_2^2 }}.
\end{equation}
The conditional average \ac{SNR} for the fading model described in (\ref{eq:12}) will be
\begin{equation}
\label{eq:17}
\begin{split}
  & \bar \gamma _{u_1 ,u_2 }  = \left( {E_s /N_0 } \right)\left( {u_1 V_1^2  + u_2 V_2^2 + 2\sigma ^2 } \right)  \cr 
  & \quad \quad  = \left( {E_s /N_0 } \right)2\sigma ^2 \left( {1 + K_{u_1 ,u_2 } } \right). \cr
\end{split}
\end{equation}
The \ac{MGF} of the TWDP fading model was shown in \cite {Rao2015} to be given in closed-form as
\begin{equation}
\label{eq:18}
\begin{split}
   \quad M_{\gamma _{_{u_1 ,u_2 } } } \left( s \right) &= {{1 + K_{u_1 ,u_2 } } \over {1 + K_{u_1 ,u_2 }  - \bar \gamma _{_{u_1 ,u_2 } } s}}  \cr 
  &  \times \exp \left( {{{K_{u_1 ,u_2 } \bar \gamma _{u_1 ,u_2 } s} \over {1 + K_{u_1 ,u_2 }  - \bar \gamma _{u_1 ,u_2 } s}}} \right) \cr 
  &  \times
	 I_0 \left( {{{\Delta _{u_1 ,u_2 } K_{u_1 ,u_2 } \bar \gamma _{u_1 ,u_2 } s} \over {1 + K_{u_1 ,u_2 }  - \bar \gamma _{u_1 ,u_2 } s}}} \right), \cr
\end{split}
\end{equation}
where $I_0(\cdot)$ is the zero-order modified Bessel function of the first kind \cite[p. 375 (9.6.12)]{Abramowitz70}. Note that, from (\ref{eq:05}) and (\ref{eq:17}), we can write 
\begin{equation}
\label{eq:19}
{{1 + K_{u_1 ,u_2 } } \over {\bar \gamma _{u_1 ,u_2 } }} = {1 \over {\left( {E_s /N_0 } \right)2\sigma ^2 }} = {{1 + K} \over {\bar \gamma }},
\end{equation}
and thus, we have
\begin{equation}
\label{eq:22}
\begin{split}
  & M_{\gamma _{_{u_1 ,u_2 } } } \left( s \right) = {{{{1 + K_{u_1 ,u_2 } } \over {\bar \gamma _{u_1 ,u_2 } }}} \over {{{1 + K_{u_1 ,u_2 } } \over {\bar \gamma _{u_1 ,u_2 } }} - s}}\exp \left( {{{K_{u_1 ,u_2 } s} \over {{{1 + K_{u_1 ,u_2 } } \over {\bar \gamma _{u_1 ,u_2 } }} - s}}} \right)  \cr 
  & \quad \quad \quad \quad \quad  \times I_0 \left( {\Delta _{u_1 ,u_2 } {{K_{u_1 ,u_2 } s} \over {{{1 + K_{u_1 ,u_2 } } \over {\bar \gamma _{u_1 ,u_2 } }} - s}}} \right)  \cr 
  & \quad \quad \quad  = {{{{1 + K} \over {\bar \gamma }}} \over {{{1 + K} \over {\bar \gamma }} - s}}\exp \left( {{{K_{u_1 ,u_2 } s} \over {{{1 + K} \over {\bar \gamma }} - s}}} \right)  \cr 
  & \quad \quad \quad \quad \quad  \times I_0 \left( {\Delta _{u_1 ,u_2 } {{K_{u_1 ,u_2 } s} \over {{{1 + K} \over {\bar \gamma }} - s}}} \right)   \cr 
  & \quad \quad \quad  = {{1 + K} \over {1 + K - \bar \gamma s}}\exp \left( {{{K_{u_1 ,u_2 } \bar \gamma s} \over {1 + K - \bar \gamma s}}} \right)  \cr 
  & \quad \quad \quad \quad \quad  \times I_0 \left( {\Delta _{u_1 ,u_2 } {{K_{u_1 ,u_2 } \bar \gamma s} \over {1 + K - \bar \gamma s}}} \right)  \cr 
  & \quad  = \mathcal{B}(s)\exp \left( {K_{u_1 ,u_2 } \mathcal{A}(s)} \right)I_0 \left( {\Delta _{u_1 ,u_2 } K_{u_1 ,u_2 } \mathcal{A}(s)} \right), \cr
\end{split}
\end{equation}
where we have defined
\begin{equation}
\label{eq:23}
\mathcal{A}\left( s \right) \triangleq \frac{{\bar \gamma s}}
{{1 + K - \bar \gamma s}},
\quad \quad
\mathcal{B}\left( s \right) \triangleq \frac{{1 + K}}
{{1 + K - \bar \gamma s}}.
\end{equation}
Considering that
\begin{equation}
\label{eq:23a}
\begin{split}
  & \Delta _{u_1 ,u_2 } K_{u_1 ,u_2 }  = {{2\sqrt {u_1 u_2 } V_1 V_2 } \over {u_1 V_1^2  + u_2 V_2^2 }}{{u_1 V_1^2  + u_2 V_2^2 } \over {2\sigma ^2 }}  \cr 
  & \quad \quad \quad \quad  = \sqrt {u_1 u_2 }{{ V_1 V_2 } \over {\sigma ^2 }}, \cr
\end{split}
\end{equation}
and using (\ref{eq:13}), the conditional \ac{MGF} can be written as
\begin{equation}
\label{eq:24}
\begin{split}
  & M_{\gamma _{_{u_1 ,u_2 } } } \left( s \right) = \mathcal{B}(s)\exp \left( {u_1 {{V_1^2 } \over {2\sigma ^2 }}\mathcal{A}(s)} \right)\exp \left( {u_2 {{V_2^2 } \over {2\sigma ^2 }}\mathcal{A}(s)} \right)  \cr 
  & \quad \quad \quad \quad \quad  \times I_0 \left( {\sqrt {u_1 u_2 } {{V_1 V_2 } \over {\sigma ^2 }}\mathcal{A}(s)} \right). \cr
\end{split}
\end{equation}
The \ac{MGF} of the \ac{SNR} of the \ac{IFTR} model can be obtained by averaging (\ref{eq:24}) over all possible realizations of the random variables $\zeta_1$ and $\zeta_2$, i.e. 
\begin{equation}
\label{eq:26}
\begin{split}
M_\gamma  \left( s \right) = \int_0^\infty  {\int_0^\infty  {M_{\gamma _{_{u_1 ,u_2 } } } \left( s \right)} f_{\zeta _1 } \left( {u_1 } \right)f_{\zeta _2 } \left( {u_2 } \right)du_1 du_2 },
\end{split}
\end{equation}
where $f_{\zeta _i } \left( {\cdot } \right)$ is given in (\ref{eq:02}).
The double integral in  (\ref{eq:26}) can be solved in closed-form by iteratively integrating with respect to variables $u_1$ and $u_2$. Thus, we can write
\begin{equation}
\label{eq:27}
\begin{split}
 M_\gamma  \left( s \right) = \mathcal{B}\left( s \right)\frac{{m_1^{m_1 } }}{{\Gamma \left( {m_1 } \right)}}\frac{{m_2^{m_2 } }}{{\Gamma \left( {m_2 } \right)}}\int_0^\infty  {u_2^{m_2  - 1} }  \\ 
 \quad \quad  \times \exp \left[ { - u_2 \left( {m_2  - \frac{{V_2^2 }}{{2\sigma ^2 }}\mathcal{A}(s)} \right)} \right]{\rm{\mathcal{I}(}}u_2 {\rm{)}}du_2, 
\end{split}
\end{equation}
where we have defined 
\begin{equation}
\label{eq:28}
\begin{split}
  & \mathcal{I}(u_2)  \triangleq \int_0^\infty  {u_1^{m_1 - 1} } \exp \left[ { - u_1 \left( {m_1  - \frac{{V_1^2 }}
{{2\sigma ^2 }}\mathcal{A}(s)} \right)} \right]  \cr 
  & \quad \quad \quad  \times I_0 \left( {\sqrt {u_1 u_2 } \frac{{V_1 V_2 }}
{{\sigma ^2 }}\mathcal{A}(s)} \right)du_1 , \cr 
\end{split}
\end{equation}
which can be calculated, using \cite[eqs. (6.643.2) and (9.220.2)]{Gradstein2007}, as
\begin{equation}
\label{eq:29}
\begin{split}
  & \mathcal{I}(u_2 {\text{) = }}\frac{{\Gamma \left( {m_1 } \right)}}
{{\left[ {m_1  - \frac{{V_1^2 }}
{{2\sigma ^2 }}\mathcal{A}(s)} \right]^{m_1 } }}  \cr 
  & \quad \quad \quad  \times \,_1 F_1 \left( {m_1 ;1;u_2 \frac{{\frac{{V_1^2 }}
{{2\sigma ^2 }}\frac{{V_2^2 }}
{{2\sigma ^2 }}\mathcal{A}^2 (s)}}
{{m_1  - \frac{{V_1^2 }}
{{2\sigma ^2 }}\mathcal{A}(s)}}} \right), \cr
\end{split}
\end{equation}
where ${}_1F_{1}(\cdot)$ is the Kummer confluent hypergeometric function \cite[p. 504 (13.1.2)]{Abramowitz70}. Introducing (\ref{eq:29}) into (\ref{eq:27}) we obtain, with the help of \cite[eq. (7.621.4)]{Gradstein2007},
\begin{equation}
\label{eq:30}
\begin{split}
  & M_\gamma  \left( s \right) = \mathcal{B}\left( s \right)\frac{{m_1^{m_1 } }}
{{\left[ {m_1  - \frac{{V_1^2 }}
{{2\sigma ^2 }}\mathcal{A}(s)} \right]^{m_1 } }}\frac{{m_2^{m_2 } }}
{{\left[ {m_2  - \frac{{V_2^2 }}
{{2\sigma ^2 }}\mathcal{A}(s)} \right]^{m_2 } }}  \cr 
  & \  \times   \, _2 F_1 \left( {m_2 ,m_1 ;1;\frac{{\frac{{V_1^2 }}
{{2\sigma ^2 }}\frac{{V_2^2 }}
{{2\sigma ^2 }}\mathcal{A}^2 (s)}}
{{\left[ {m_1  - \frac{{V_1^2 }}
{{2\sigma ^2 }}\mathcal{A}(s)} \right]\left[ {m_2  - \frac{{V_2^2 }}
{{2\sigma ^2 }}\mathcal{A}(s)} \right]}}} \right). \cr
\end{split}
\end{equation}
Considering now that, from (\ref{eq:03}) and (\ref{eq:04}),
\begin{equation}
\label{eq:31}
\begin{split}
\frac{{V_1^2 }}
{{2\sigma ^2 }} = \frac{K}
{2}\left( {1 + \sqrt {1 - \Delta ^2 } } \right),\ \frac{{V_2^2 }}
{{2\sigma ^2 }} = \frac{K}
{2}\left( {1 - \sqrt {1 - \Delta ^2 } } \right),
\end{split}
\end{equation}
we finally obtain (\ref{eq:06}). 

\section{Proof of Corollary 1}
\label{App2}

For $m\in\mathbb{Z}^+$ the Kummer hypergeometric function can be expressed in terms of the Laguerre polynomials by using \cite[eq. 24]{erdelyi1940}, and from the well-known Kummer transformation we have
\begin{align}
\label{eq:a07}
{}_1{F}_1(m,1;z)=e^z \sum_{n=0}^{m-1}\binom{m-1}{n}\frac{z^n}{n!}.
\end{align}
Thus, for $m_1\in\mathbb{Z}^+$, the function $\mathcal{I}(\cdot )$ given in  (\ref{eq:29}) can be written as 
\begin{equation}
\label{eq:a08}
\begin{split}
  & \mathcal{I}(u_2 ) = \frac{{\Gamma \left( {m_1 } \right)}}
{{\left[ {m_1  - \frac{{V_1^2 }}
{{2\sigma ^2 }}\mathcal{A}(s)} \right]^{m_1 } }}\exp \left( {u_2 \frac{{\frac{{V_1^2 }}
{{2\sigma ^2 }}\frac{{V_2^2 }}
{{2\sigma ^2 }}\mathcal{A}^2 (s)}}
{{m_1  - \frac{{V_1^2 }}
{{2\sigma ^2 }}\mathcal{A}(s)}}} \right)  \cr 
  & \quad \quad  \times \sum\limits_{n = 0}^{m_1  - 1} {\frac{1}
{{n!}}\binom{m_1-1}{n} \left( {u_2 \frac{{\frac{{V_1^2 }}
{{2\sigma ^2 }}\frac{{V_2^2 }}
{{2\sigma ^2 }}\mathcal{A}^2 (s)}}
{{m_1  - \frac{{V_1^2 }}
{{2\sigma ^2 }}\mathcal{A}(s)}}} \right)^n .\quad }  \cr
\end{split}
\end{equation}
Introducing (\ref{eq:a08}) into (\ref{eq:27}) the resulting integral can be solved in closed-form, yielding (\ref{eq:100}) after some manipulation.

\section{Proof of Lemma 2}
\label{App3}

The \ac{MGF} of $\gamma$ given in (\ref{eq:100}) can be factorized as
\begin{equation}
\label{eq:38}
\begin{split}
  & M_\gamma  \left( s \right) =  - \left( {1 + K} \right)m_1^{m_1 } m_2^{m_2 } a_1^{m_1  - m_2 } \sum\limits_{n = 0}^{m_1  - 1} {\frac{1}
{{n!}}}   \cr 
  & \quad  \times \binom{m_1-1}{n}\frac{{\Gamma \left( {m_2  + n} \right)}}
{{\Gamma \left( {m_2 } \right)}}\left( {\frac{{K\Delta }}
{2}} \right)^2 \frac{1}
{{a_2^{m_2  + n} }}\frac{1}
{{\bar \gamma s}}  \cr 
  & \quad  \times \left( {1 - \frac{{1 + K}}
{{\bar \gamma s}}} \right)^{m_1  - n - 1} \left( {1 - \frac{{m_1 (1 + K)}}
{{a_1 \bar \gamma s}}} \right)_{}^{m_2  - m_1 }   \cr 
  & \quad  \times \left( {1 - \frac{{m_1 m_2 (1 + K)}}
{{a_2 \bar \gamma s}}} \right)^{ - (m_2  + n)} , \cr
\end{split}
\end{equation}
where we have defined
\begin{equation}
\label{eq:39}
\begin{split}
  & a_1  \triangleq m_1  + \frac{K}
{2}\left( {1 + \sqrt {1 - \Delta ^2 } } \right)  \cr 
  & a_2  \triangleq m_1 \frac{K}
{2}\left( {1 - \sqrt {1 - \Delta ^2 } } \right)  \cr 
  & \quad \quad \quad  + m_2 \frac{K}
{2}\left( {1 + \sqrt {1 - \Delta ^2 } } \right) + m_1 m_2 . \cr
\end{split}
\end{equation}

Taking into account that the \ac{PDF} is related to the \ac{MGF} by the inverse Laplace transform, i.e., 
$f_{\gamma}(x)=\L^{-1}[M_{\gamma}(-s);x]$,  (\ref{eq:36}) follows from (\ref{eq:38}) and the
Laplace transform pair given in \cite[p. 290, (55)]{Srivastava1985}. On the other hand, (\ref{eq:37}) is obtained analogously by considering that  $F_{\gamma}(x)=\L^{-1}[M_{\gamma}(-s)/s;x]$.

\bibliographystyle{IEEEtran}
\IEEEtriggeratref{5}
\bibliography{bibjavi}
\end{document}